# `A15 Phase $Ta_3Sb$ Thin Films: Direct Synthesis, Charge Transport and Spin-Orbit Torque

J. S. Jiang[1*], Qianheng Du[1], Yi Li[1], Ulrich Welp[1], Ramakanta Chapai[1], Hanu Arava[1], Yuzi Liu[2], Yue Li[1], John Pearson[1], Ralu Divan[2], Anand Bhattacharya[1], and Hyowon Park[1,3]

1. Materials Science Division, Argonne National Laboratory, Lemont, IL 60493

2. Nanoscience and Technology Division, Argonne National Laboratory, Lemont, IL 60493

3. Department of Physics, University of Illinois at Chicago, Chicago, IL 60607

## Abstract

$Ta_3Sb$ is one of the A15 compounds that have been predicted to have giant spin Hall conductivities due to the gapped Dirac-like band crossings in their electronic structures. We use co-sputtering to directly synthesize thin films of $Ta_3Sb$ and identify a large window of Ta:Sb flux ratio that permits the formation of single-phase A15 structure. These sputtered films have an actual Ta:Sb atomic ratio of 4:1 as determined from Rutherford backscattering spectrometry. Their high resistivity, at the Mott–Ioffe–Regel limit, suggests that the electron mean free path is comparable to interatomic distances. From harmonic Hall and spin-torque ferromagnetic resonance measurements, the intrinsic spin Hall conductivity of thin film $Ta_3Sb$ is estimated to be in the range of -526 to -1230 $(\hbar/e)\,S/cm$ at 300 K, lower in magnitude than the predicted value of $-1400\,(\hbar/e)\,S/cm$. First-principles calculations of the electronic structure show that the discrepancy is consistent with an increase of the Fermi level due to the non-ideal stoichiometry needed to stabilize the A15 structure.

[*] Corresponding author: jiang@anl.gov

## I. INTRODUCTION

Current-induced spin-orbit torques (SOTs) have become a promising tool in spintronic applications to electrically manipulate magnetic moments with a lateral current [1-3]. Of significant interest in implementing SOT-based technologies is the development of materials capable of efficient charge-spin conversion. The efficiency of converting between a charge current $j_c$ and spin current $j_s$ is characterized by the SOT efficiency $\xi = (2e/\hbar)(j_s/j_c) = T_{int}\theta_{SH} \equiv T_{int}\sigma_{SH}/\sigma$, where $T_{int}$, $\theta_{SH}$, $\sigma_{SH}$ and $\sigma$ are the interfacial spin transparency, spin Hall angle, spin Hall conductivity and electrical conductivity, respectively. Thus, a large $\sigma_{SH}$ can lead to a high $\xi$. Conventional heavy metals such as Pt and $\beta$-W have in their electronic structures degenerate bands that are split by spin-orbit coupling (SOC). This leads to non-vanishing Berry curvature which, when integrated over the occupied bands, gives rise to large $\sigma_{SH}$ [4]. Derunova *et al.* [5] point out that a strategy to search for or design new materials with high $\sigma_{SH}$ is to maximize Berry curvature by having the Fermi level $E_F$ inside as many small-gapped band crossings as possible. They note that for many intermetallic compounds with the A15 structure, due to their crystalline symmetry (space group 223, $Pm\bar{3}n$), the calculated band structures contain Dirac crossings near $E_F$ if SOC is not considered. These crossings become gapped when SOC is included, leading to giant values of $\sigma_{SH}$. It is predicted that the A15 phase $W_3Ta$ has a very large $\sigma_{SH}$ of $-2250\ (\hbar/e)\ S/cm$, and enhancements in spin Hall ratio [6] and effective spin Hall angle [7] have been reported when $\beta$-W is doped with Ta. Remarkably, the reported enhancements (9% vs. 100%) are vastly different despite similar levels (8.5 vs. 11 atomic %) of Ta-doping.

The A15 superconductor $Ta_3Sb$ has also been predicted to have a large $\sigma_{SH} = -1400\ (\hbar/e)\ S/cm$ with an SOC-gapped eight-fold degenerate Dirac point near $E_F$. Moreover, there exist in $Ta_3Sb$ near $E_F$ topologically non-trivial surface states that may become proximitized by the bulk superconductivity, making $Ta_3Sb$ a potential candidate to explore topological superconductivity [5, 8, 9]. Because of the large difference in the melting points of Ta and Sb and the high vapor pressure of Sb, direct synthesis of $Ta_3Sb$ has not been possible. Chapai *et al.* [10] synthesized phase-pure powders of A15 $Ta_3Sb$ from the Sb-rich precursor $TaSb_2$ via a lengthy 14-day thermal decomposition process. The superconducting properties and specific heat of $Ta_3Sb$ were studied using pellets pressed from the powders. However, due to the porous nature of the pressed pellets, it was not possible to fabricate the device structures that are necessary for characterizing spin-orbit effects.

In this work we directly synthesize thin films of $Ta_3Sb$ from elemental sources using the co-sputtering process. We identify a large window of processing conditions that can stabilize single-phase A15 $Ta_3Sb$, and analyze the structure and charge transport characteristics of the thin films. The chemical composition of thin film $Ta_3Sb$ deviates from the ideal stoichiometry but agrees with that of $Ta_3Sb$ bulk crystals [10] and is as expected from the Ta-Sb equilibrium phase diagram [11]. From harmonic Hall and spin-torque ferromagnetic resonance measurements, we estimate a range for the spin Hall conductivity of $Ta_3Sb$, which is lower in magnitude than the theoretical prediction. We perform density functional theory calculations for the electronic structure and explain that the discrepancy in spin Hall conductivity is due to the non-ideal stoichiometry of the actual A15-phase $Ta_3Sb$. We highlight the importance of the location of $E_F$ of realizable A15 materials and discuss possible avenues to optimize their spin Hall conductivity.

## II. THIN FILM SYNTHESIS AND STRUCTURAL CHARACTERIZATION

The $Ta_3Sb$ films were grown by magnetron sputtering from elemental Ta and Sb targets onto silicon (100) wafers with a native oxide layer. Both doped (resistivity $10\ \Omega \cdot cm$) and intrinsic (resistivity > 20 k$\Omega \cdot cm$) wafers were used. The choice of substrate had no effect on film growth. The base pressure of the deposition system before growth was better than $3 \times 10^{-9}$ Torr. The working pressure of the Ar sputter gas was maintained at 3 mTorr. The source fluxes were set with a calibrated quartz microbalance before deposition, and were measured after deposition to ensure there had been no significant drift. Because of the high vapor pressure of Sb, an over-supply of Sb flux was necessary to compensate for losses during film growth. We varied the substrate temperature and the Ta:Sb flux ratio to search for the optimal growth conditions, guided by crystal structure and phase analyses of X-ray diffraction (XRD) measurements. Fig. 1 shows the XRD patterns of Ta-Sb films grown at a substrate temperature of 650°C with the Ta:Sb flux ratio ranging from 3:1 to 1:2. A threshold Ta:Sb flux ratio of approximately 2:1 was needed for the formation of single-phase A15 $Ta_3Sb$ films. Further decreasing the flux ratio by up to a factor of 4 did not produce any detectable impurity phase or change in the lattice constant of the A15 structure. This result points to a film growth process that is similar to the "phase-locking" approach used for preparing A15 $Nb_3Sn$ films [12]: at suitably controlled substrate temperatures, the film composition always settles to the boundary of the A15 phase while the surplus high-vapor-pressure component (Sn or Sb) simply re-evaporates, thereby allowing for the direct synthesis of homogeneous A15 films without the need to precisely control the source fluxes. Subsequently, we fixed the substrate temperature at 650°C and the Ta:Sb flux ratio at 3:2 for all depositions. The effective growth rate calculated from deposition times and the film thicknesses as measured with X-ray reflectivity (XRR) was approximately 2 nm/min. The films were cooled to room temperature at a rate of 20°C/min. Except for the single-layer $Ta_3Sb$ films prepared for composition analyses, all films were capped with 5 nm of MgO before being taken out of vacuum.

Shown in Fig. 1(b) is a representative XRD $2\theta - \omega$ scan of an 85 nm-thick $Ta_3Sb$ film measured using Cu $K_\alpha$ radiation. The film is polycrystalline. To make the higher-index diffraction peaks in the $2\theta$ range of 60°-75° more discernable, we intentionally offset $\omega$ by 2.5° from $\theta$ so as to suppress the strong Si (200) reflection. All the observed diffraction peaks can be indexed to those of $Ta_3Sb$ with the A15 structure. The lattice constant $a$ is determined from the peak positions to be $0.5272 \pm 0.0004$ nm, which is very similar to the value reported for bulk crystals of $Ta_3Sb$ [10]. Compared with the $Ta_3Sb$ powder diffraction pattern where the (200) peak has less than half the intensity of the (211) peak, the (200) diffraction peak of the sputtered film is nearly twice as strong as the (211) peak, indicating some (200) texturing in the growth direction. From the full width at half maximum of the diffraction peaks we estimate the average coherence length of the $Ta_3Sb$ crystallites to be about 40 nm in the growth direction. The XRR measurement on an identically grown $Ta_3Sb$ film is shown in the inset of Fig. 1(b). Fitting the data yields a layer thickness of 84.8 nm and a $Ta_3Sb$/MgO interfacial roughness of 1.2 nm.

Although the XRD pattern and the lattice constant of the sputtered films match those of the A15 $Ta_3Sb$, their actual chemical composition deviates from that of the ideal stoichiometry. We analyzed the composition of our films using Rutherford backscattering spectrometry (RBS). As shown in Fig. 2, curve-fitting the RBS spectrum taken on a 5.4 nm-thick sputtered $Ta_3Sb$ film yields areal densities of $(43.23 \pm 0.01) \times 10^{15}$ and $(11.04 \pm 0.01) \times 10^{15}$ $atoms/cm^2$ for Ta and Sb, respectively. This corresponds to a Ta:Sb atomic ratio of 4:1, indicating that the film is Sb-deficient. Similar Sb deficiency has been also reported in bulk crystals, and is consistent with the Ta-Sb phase diagram [11], where the A15 phase exists in a narrow homogeneity region that is closer in composition to $Ta_4Sb$ and is characterized by a random mixing of Ta and Sb atoms at the Sb site [10]. As is typically done for A15 compounds that deviate from their ideal stoichiometry, we refer to the sputtered films by their prototypical formula as $Ta_3Sb$ throughout.

To assess the microstructure and compositional uniformity of the sputtered films, we carried out cross-sectional imaging using transmission electron microscopy (TEM) and energy-dispersive X-ray spectroscopy (EDS). The nano-beam electron diffraction was performed in a JEOL 2100f electron microscope, and the EDS mapping was performed in a FEI Talos F200X TEM/STEM. The dark-field (DF) cross-sectional TEM image and nanobeam electron diffraction patterns in Fig. 3(a) demonstrate the microstructure of an 85 nm $Ta_3Sb$ film. In the DF image, a columnar microstructure is visible where grains are differently oriented crystallites, as shown by the indexed nanobeam electron diffraction patterns for select areas. The columnar grains are about 10-20 nm wide and only extend to about half the thickness of the film, consistent with the coherence length estimated from the XRD peak widths. In Fig. 3(b), the High-Angle Annular Dark Field (HAADF) image and the corresponding EDS maps of Ta and Sb are displayed. The EDS maps show uniform distribution of the elements. Since $Ta_5Sb_4$ and $TaSb_2$ are the only other possible Ta-Sb alloy phases, and since both are line compounds, the lack of modulation in EDS intensity at the length scale of grain sizes confirms that our sputtered film is single-phase $Ta_3Sb$.

## III. CHARGE TRANSPORT PROPERTIES

For charge transport properties, films grown on high-resistivity undoped Si were patterned into 6-terminal Hall bar devices using standard optical lithography and argon ion milling. The current channel was nominally 10 $\mu m$ wide, and 40 $\mu m$ long between the voltage terminals. The width of the voltage terminals was 2 $\mu m$. The transport measurements were carried out at temperatures between 50 mK and 300 K in magnetic fields up to 14 T using a Quantum Design PPMS system with a dilution refrigerator insert.

Figure 4(a) shows the electrical resistivity of an 85 nm $Ta_3Sb$ thin film as a function of temperature between 300 K and 5 K. The resistivity has a negative temperature coefficient (TCR), increasing from 186 $\mu\Omega \cdot cm$ at 300 K to 192 $\mu\Omega \cdot cm$ at 5 K. Similarly, the resistivity of a 9.5 nm $Ta_3Sb$ film increases from 224 $\mu\Omega \cdot cm$ at 300 K to 237 $\mu\Omega \cdot cm$ at 5 K. The very large values of resistivity, comparable to the saturation resistivity of metallic elements and alloys [13], implies that the $Ta_3Sb$ thin films are at the Mott–Ioffe–Regel limit where the electron mean free path is limited to interatomic distances. The mixing of Ta and Sb atoms at the Sb sites arising from the non-ideal stoichiometry could provide the needed strong disorder. The negative TCR could be due to Anderson localization [14], a strong coupling polaronic mechanism [15] or, as originally proposed by Mooij, an increase in the number of charge carriers with temperature [16]. The carrier density in the $Ta_3Sb$ thin films is indeed temperature dependent. Shown in Fig. 4(b) are the Hall resistance $R_{xy}$ of the 85 nm $Ta_3Sb$ film as a function of magnetic field measured at various temperatures. $R_{xy}$ is linear with fields as high as 14 T, and the slopes are positive, indicating the dominant charge carriers are hole-type. The carrier density is comparable to that of typical metals, increasing slightly with temperature from $4.8 \times 10^{28}\ m^{-3}$ at 2 K to $5.4 \times 10^{28}\ m^{-3}$ at 200 K (see Fig. 4(b) inset). This temperature-dependent carrier density might explain the negative TCR in the $Ta_3Sb$ thin films. It is worth nothing that some metals such as aluminum also have positive Hall coefficients due to the peculiar shapes of their Fermi surfaces.

In Fig. 5, the resistive superconducting transitions of the $Ta_3Sb$ film measured in various out-of-plane magnetic fields are shown. The superconducting transition temperature $T_c$, defined as the midpoint of the resistive transition, is 0.887 K at zero field. $T_c$ is progressively suppressed with increasing magnetic field. The upper critical field $H_{c2}$ as a function of temperature is plotted in the inset of Fig. 5. $H_{c2}$ varies linearly with temperature above 0.3 K but does not exhibit the downward curvature that was observed in $Ta_3Sb$ bulk crystals, indicating that superconductivity in the $Ta_3Sb$ thin films is not confined by grain boundaries and/or voids as it is in the bulk crystals, even though the in-plane extent (10-20 nm) of the crystallites in the thin films is smaller than the grain size (~45 nm) of the bulk crystals. Extrapolating $H_{c2}$ to zero temperature yields the Ginzburg-Landau upper critical field of $H_{c2}^{GL} = 1.25\, T$, which is slightly higher than that of the bulk crystals, consistent with more disorder being present in the $Ta_3Sb$ thin films.

## IV. SPIN-ORBIT TORQUE EFFECTS

To quantify the spin Hall angle in the A15 $Ta_3Sb$ thin films, we employed two complementary techniques — harmonic Hall [17] and spin-torque ferromagnetic resonance (ST-FMR) [18] — to measure the current-induced spin-orbit torques in bilayer structures consisting of a $Ta_3Sb$ layer and a soft ferromagnetic layer of $Co_{40}Fe_{40}B_{20}$ (CFB) or $Ni_{81}Fe_{19}$ (Py). The CFB and Py layers were deposited at room temperature after cooling the $Ta_3Sb$ layer without breaking vacuum. The bilayers were then capped with 5 nm of MgO before being taken out of vacuum for lithographic patterning. For harmonic Hall measurements, the bilayers were patterned into the same Hall bar structure as that used for charge transport measurements. For ST-FMR measurements, rectangular microstrips $20\mu m \times 50\mu m$ in size were patterned from the bilayer films and were integrated with gold co-planar waveguides for radio frequency (RF) excitation [19, 20].

In a harmonic Hall measurement, as schematically shown in Fig. 6(a), the spin current induced by an AC current $I_{AC}$ exerts oscillating damping-like (DL) and field-like (FL) effective

SOT fields $H_{DL}$ and $H_{FL}$ on the magnetization of the ferromagnet. The oscillation of the magnetization around its equilibrium position gives rise to a second harmonic component in the Hall voltage via the anomalous Hall and planar Hall effects. For in-plane magnetized systems, when an out-of-plane magnetic field $H$ is swept, the second harmonic component of the Hall voltage is given by [17]:

$$V_{2\omega} = \frac{I_{AC}R_{AHE}\sin\theta_0\cos\varphi_H}{2(H_K\cos 2\theta_0 + H\cos(\theta_H - \theta_0))} H_{DL} - \frac{I_{AC}R_{PHE}\sin^2\theta_0\cos\varphi_H}{H\sin\theta_H}(H_{FL} + H_{Oe}). \quad (1)$$

Here $R_{AHE}$ and $R_{PHE}$ are the anomalous Hall and planar Hall resistances, respectively. $\theta_0$ is the polar angle of the magnetization and can be obtained from the first harmonic Hall voltage $V_\omega = I_{AC}R_{AHE}\cos\theta_0$. $H_{Oe}$ is the Oersted field produced by the current, and $H_K$ is the effective out-of-plane anisotropy field. $\theta_H$ and $\varphi_H$ are the polar and azimuthal angles of the direction of the applied magnetic field, respectively. A notable feature of Eqn. (1) is that while the second term, the $H_{FL}$ contribution, has an inverse $H$ dependence, the $H_{DL}$ contribution in the first term peaks when $H$ is comparable to $-H_K$ , i.e. when the applied field compensates the shape anisotropy and the DL SOT becomes more effective at driving the magnetization oscillation. Importantly, an out-of-plane temperature gradient generally exists in harmonic Hall measurements because of significant Joule heating, and while harmonic Hall measurements with a rotating in-plane magnetic field need to account for thermoelectric effects [21, 22], the measurement of $V_{2\omega}$ when the magnetic field is applied out-of-plane is not susceptible to the out-of-plane temperature gradient.

Figure 6(b) shows the first harmonic ($V_\omega$) and second harmonic ($V_{2\omega}$) Hall voltages of a $Ta_3Sb$(9.5nm)/CFB(1nm) bilayer measured as a function of the applied out-of-plane magnetic field $H$ at various temperatures. The offsets due to Hall lead misalignment and the linear backgrounds due to the ordinary Hall effect have been removed from the $V_\omega$ data, whereas no processing has been done on the $V_{2\omega}$ data. The sample has in-plane magnetization. The applied AC current is $I_{AC} = 3\ mA$. The applied field is intentionally tilted by 5° away from the surface normal to avoid domain formation (*i.e.* $\theta_H = 5°$ when $\varphi_H = \pi$ or $\theta_H = 85°$ when $\varphi_H = 0$). With an increasing out-of-plane field, the $V_\omega$ curves show the magnetization being pulled out of plane, reaching saturation when the applied field overcomes the effective in-plane anisotropy. From the $V_\omega$ curves, $R_{AHE}$ and $H_K$ can be determined. $H_K$ is the field at which a line extrapolated from the initial rise of $V_\omega$ intersects the saturation. $|H_K|$ at 300K is 0.17 T which is significantly lower than the demagnetizing field $4\pi M_s = 1.5$ T for CFB because of the presence of a perpendicular magnetic anisotropy at the CFB/MgO interface. Here $M_s$ is the saturation magnetization of a single CFB layer measured using vibrating sample magnetometry. We also separately measured $V_\omega$ while rotating the applied field in the sample plane as shown in Fig. 6(c) and determined $R_{PHE}$. In Fig. 6(b), the $V_{2\omega}$ curves exhibit bumps that clearly signify the presence of DL SOT and, for $T \leq 200\ K$, asymptote towards zero. For higher temperatures, such as the $T = 300\ K$ case shown, the $V_{2\omega}$ curves have linear-in-H backgrounds which we attribute to the ordinary Nernst effect arising from *in-plane* temperature gradients due to lateral asymmetries of the Hall bar structure. With $R_{AHE}$, $R_{PHE}$ and $\theta_0$ determined, we extracted $H_K$, $H_{DL}$ and $H_{FL}$ by fitting the non-hysteretic portions of the $V_{2\omega}$ curves with Eqn. (1). The $H_K$ values from the curve fits agree well with the saturation fields obtained from the $V_\omega$ curves.

The efficiencies at which the current generates DL and FL SOTs ($\xi_{DL}$ and $\xi_{FL}$) are given by equating the transferred angular momentum from the spin current with the effect of $H_{DL}$ and $H_{FL}$ on the magnetization [23]:

$$\xi_{DL(FL)} = \frac{2e\mu_0 M_s t H_{DL(FL)}}{\hbar j_c}, \qquad (2)$$

where $t$ is the thickness of the ferromagnetic CFB layer. Using $M_s = 1.19\ MA/m$, we determined $\xi_{DL}$ and $\xi_{FL}$ of the $Ta_3Sb$ films. Fig. 6(d) summarizes $\xi_{DL}$ and $\xi_{FL}$ at various temperatures. $\xi_{DL}$ and $\xi_{FL}$ are opposite in sign. $\xi_{DL}$ decreases from -0.087 at 20 K to -0.116 at 300K, while $\xi_{FL}$ increases from 0.013 at 20 K to 0.019 at 300 K.

On the other hand, in an ST-FMR measurement, the RF current produces high frequency SOTs which excite resonant oscillation of the magnetization at the resonant frequency $f_{res}$ and the resonant field $H_{res}$ [18]. We applied RF excitations to $Ta_3Sb$/Py bilayer structures at frequencies from 4 GHz to 10 GHz and measured $V_{mix}$ while sweeping the applied magnetic field. The RF power was kept constant at 11.5 dBm. The mixing of the RF current and current-induced RF anisotropic magnetoresistance gives rise to a DC voltage $V_{mix}$, which consists of symmetric and antisymmetric Lorentzian components with magnitudes of $S$ and $A$, respectively:

$$V_{mix} = S\frac{\Delta H^2}{\Delta H^2+(H-H_{res})^2} + A\frac{\Delta H(H-H_{res})}{\Delta H^2+(H-H_{res})^2}. \quad (3)$$

Here $\Delta H$ is half of the linewidth. $S$ is proportional to $H_{DL}$ and $A$ is proportional to $H_{FL} + H_{Oe}$. The DL SOT efficiency is thus given by [24]:

$$\xi_{DL} = \frac{S}{A}\frac{e\mu_0 M_s t d}{\hbar}\sqrt{1+\frac{4\pi M_{eff}}{H_{res}}}\left(1+\frac{\hbar\xi_{FL}}{e\mu_0 M_s t d}\right). \quad (4)$$

Here $t$ and $d$ are the thicknesses of the Py and $Ta_3Sb$ layers, respectively, and $M_{eff}$ is the effective magnetization of Py.

Figure 7(a) shows typical ST-FMR spectra for a $Ta_3Sb$(9.5nm)/Py(5nm) bilayer measured at 300K. The values of $H_{res}$, $\Delta H$, $S$ and $A$ are obtained by fitting the ST-FMR spectra using Eqn. (3). As shown in Figure 7(b), by fitting the resonant frequency $f_{res}$ as a function of $H_{res}$ using the Kittel formula $f_{res} = (\gamma/2\pi)\left[H_{res}\left(H_{res} + 4\pi M_{eff}\right)\right]^{1/2}$ and $\gamma = 1.855 \times 10^{11}\ Hz/T$ [25], we obtain $M_{eff} = 697\ kA/m$. The value of $M_{eff}$ is similar to that of $M_s = 716\ kA/m$ for the Py layer determined from vibrating sample magnetometry. With ST-FMR alone, $\xi_{DL}$ and $\xi_{FL}$ cannot be independently determined without fabricating a series of bilayers with different $d$ or $t$ while maintaining consistent qualities for the layers and interfaces. Nevertheless, given that $\xi_{FL}$ is an intrinsic quantity, we can take the value of $\xi_{FL}$ determined from harmonic Hall measurements on $Ta_3Sb$/CFB. Using $\xi_{FL} = 0.019$ in Eqn. (4), we obtain the average $\xi_{DL} = -0.098$ for $Ta_3Sb$/Py, which is quite close to the $\xi_{DL} = -0.116$ obtained for $Ta_3Sb$/CFB.

The SOT efficiency $\xi$ is given by the intrinsic spin Hall angle $\theta_{SH}$ and the interfacial spin transparency $T_{int}$ via $\xi = T_{int}\theta_{SH}$. The values of $|\xi_{DL}|$ can be considered as the lower bound for $|\theta_{SH}|$ in the event of a spin-transparent interface, i.e., when $T_{int} = 1$. However, $T_{int}$ is generally less than unity at realistic interfaces due to spin mixing [24, 26, 27]:

$$T_{int} = 2g_{eff}^{\uparrow\downarrow}/[(h/e^2)G_{NM}], \tag{5}$$

where $g_{eff}^{\uparrow\downarrow}$ is the effective spin mixing conductance, $G_{NM} = \sigma/\lambda$ is the spin conductance of the $Ta_3Sb$ layer, and $\lambda$ is the spin diffusion length. Eqn. (5) is valid when the $Ta_3Sb$ layer thickness is far greater than $\lambda$ [24]. Given that the $Ta_3Sb$ layer is in the Mott–Ioffe–Regel limit, it is reasonable to assume this condition is satisfied. We evaluate $g_{eff}^{\uparrow\downarrow}$ from linewidth broadening in the ST-FMR spectra, because spin mixing enhances the Gilbert damping [28, 29]: $\Delta\alpha = \alpha - \alpha_0 = (\frac{\gamma\hbar}{4\pi M_S t_{Py}})g_{eff}^{\uparrow\downarrow}$,. Figure 7(c) shows the half linewidth $\Delta H$ extracted from the ST-FMR spectra as a function of the resonant frequency $f_{res}$. Fitting the data points with the linear relation $\Delta H = (2\pi\alpha/\gamma)f_{res} + \Delta H_0$ yields the Gilbert damping parameter $\alpha = 0.0118$. Using $\alpha_0 = 0.0090$ measured from a sputtered Py single layer [30], we obtain $g_{eff}^{\uparrow\downarrow} = 6.48 \times 10^{18}\ m^{-2}$. Since the disorder in the sputtered A15 $Ta_3Sb$ films is primarily the replacement of Sb atoms by Ta atoms at the Sb sites, we take shortest physical length scale that may correspond to spin flip scattering, i.e., the closest distance between two Sb sites ($a\sqrt{3}/2$), as the lower bound for $\lambda$ to estimate the lower bound for $T_{int}$. This yields the lower bound of 0.43 for $T_{int}$ at the $Ta_3Sb$/Py interface, and thus the upper bound for $|\theta_{SH}|$ is 0.230. Therefore, the A15 $Ta_3Sb$ has an intrinsic spin Hall angle $\theta_{SH}$ in the range of -0.098 to -0.230, and its spin Hall conductivity $\sigma_{SH}$ is in the range of -526 to $-1230(\hbar/e)\ S/cm$.

## V. ELECTRONIC STRUCTURE

We note that the wide ranges for $\theta_{SH}$ and $\sigma_{SH}$ are due to uncertainties in the interfacial spin transparency $T_{int}$. Nevertheless, even with our conservative estimate of $\lambda$, the maximum possible value for $|\sigma_{SH}|$ in the sputtered A15 $Ta_3Sb$ thin film is smaller than the theoretical prediction of $1400\ (\hbar/e)\ S/cm$. The discrepancy may be explained by the fact that the theoretical prediction is based on band structure calculations for the stoichiometric A15 $Ta_3Sb$, while our thin films of A15 $Ta_3Sb$ -- as well as bulk crystals of A15 $Ta_3Sb$ -- are stabilized with the chemical composition of Ta:Sb ~ 4:1 (or, $Ta_{3.2}Sb_{0.8}$, to emphasize the replacement of 20% Sb by Ta due to site-mixing [10]). To examine the effect of off-stoichiometry on spin Hall conductivity, we performed first-principles density functional theory (DFT) calculations to evaluate changes in the Fermi energy $E_F$ as a result of electron doping from the excess Ta atoms.

We start by calculating the band structure of the stoichiometric A15 $Ta_3Sb$ using the projector augmented wave (PAW) method as implemented in the Vienna Ab initio Simulation Package (VASP) code [31, 32]. We converge the DFT calculations using a $10 \times 10 \times 10$ k-mesh and an energy cutoff of 400 eV for the plane-wave basis. We use the Perdew-Burke-Ernzerhof (PBE) functional for the exchange and correlation energy of DFT. Figure 8(a) shows the PBE band structure calculated with SOC included. It is in excellent agreement with that of Ref. [5]. We then evaluate $E_F$ for different numbers of occupied electrons ($N_{occ}$), assuming that the band structure is shifted rigidly (the Kohn-Sham eigenvalues are fixed) and only $E_F$ is changed due to doping. We recompute the Kohn-Sham equation for each $N_{occ}$, while fixing the charge density obtained from the converged solution during iterations to impose the same band structure. For the summation of the occupied bands and k-points below $E_F$, we adopt the tetrahedron method using the $10 \times 10 \times 10$ k-mesh. The resulting change in $E_F$ ($\Delta E_F$) as a function of $N_{occ}$ per formula unit (f. u.) is shown in Fig. 8(b). $Ta_{3.2}Sb_{0.8}$ has 0.4 more valence electrons per formula unit than $Ta_3Sb$, we estimate that the extra electrons increase $E_F$ by 0.425 eV and, according to Fig. 4C of Ref. [5], would move $\sigma_{SH}$ off its peak to $\sim -900\ (\hbar/e)\ S/cm$. This reduced value falls within our estimated range for $\sigma_{SH}$ in the sputtered A15-phase $Ta_3Sb$ thin films.

## VI. DISCUSSION AND SUMMARY

That the shift of $E_F$ away from the SOC-gapped Dirac crossings leads to a reduced $\sigma_{SH}$ in the off-stoichiometric A15-phase $Ta_3Sb$ validates the strategy by Derunova *et al.* [5] of searching for high $\sigma_{SH}$ in A15 compounds due to their symmetry, in the same way that Ta-doped W exhibiting enhanced $\sigma_{SH}$ [6, 7] does. The opposite end results of these two scenarios, however, underscore the importance of the $E_F$ of the actual realizable material, since many of the A15 compounds are stable only in compositions that are different from the ideal stoichiometry [33, 34]. To maximize $\sigma_{SH}$, it might be beneficial to investigate pseudo-binary A15 alloys such as $Ta_3$(Sb, Sn), where the Sn dopant could take up the excess electrons and return $E_F$ to the optimum level. On the other hand, we note that even with the raised $E_F$ in the off-stoichiometric A15 $Ta_3Sb$, the topological surface states are still accessible [5, 8], making the A15 $Ta_3Sb$ thin films potentially useful for exploring topological superconductivity.

In summary, we have synthesized single-phase A15 $Ta_3Sb$ thin films by co-sputtering. The phase-locking growth mechanism allows for facile and direct synthesis. However, the A15 $Ta_3Sb$ is stabilized at a composition that is different from the ideal stoichiometry. This leads to the shift of the Fermi level aways from the SOC-gapped Dirac crossings where the maximum spin Hall conductivity is expected. It is therefore important to note the thermodynamic constraints of materials synthesis. A possible approach to designing giant-spin-Hall-effect A15 compounds may lie in tuning the Fermi level via pseudo-binary alloys.

## ACKNOWLEDGEMENT

We acknowledge Dr. Timothy Spila for the acquisition of RBS data. Work at Argonne National Laboratory, including film growth and structural characterization, transport measurements and theoretical calculations, was supported by the U.S. Department of Energy Office of Science, Basic Energy Sciences, Materials Science and Engineering Division. The use of the microscopy and microfabrication facilities at the Center for Nanoscale Materials was supported by the U.S. DOE, BES under Contract No. DE-AC02-06CH11357. H. Park acknowledges the computing resources provided on Bebop, a high-performance computing cluster operated by the Laboratory Computing Resource Center at Argonne National Laboratory. The ST-FMR experiments at Argonne National Laboratory was supported by the U.S. DOE, Office of Science, Basic Energy Sciences, Materials Sciences and Engineering Division under Contract No. DE-SC0022060.

**Figure captions:**

Figure 1: (a) XRD patterns of Ta-Sb films sputter-grown at various Ta:Sb flux ratios. The patterns are shifted vertically for clarity. The XRD scattering vector is aligned along the surface normal. The diffraction peaks marked with the symbol "o" are due to the silicon substrate. (b) XRD pattern of an 85 nm $Ta_3Sb$ film. Inset: XRR data and fit. The slower modulation in intensity is due to the MgO capping layer.

Figure 2: RBS data and fit of a 5.4 nm $Ta_3Sb$ film.

Figure 3: (a) Dark-field cross-sectional TEM image of an 85 nm $Ta_3Sb$ film. Nanobeam electron diffraction patterns of selected areas A and B are shown. (b) HAADF image and corresponding EDS maps of Ta and Sb showing the elemental distributions.

Figure 4: (a) Electrical resistivity of an 85 nm $Ta_3Sb$ film plotted as a function of temperature. (b) Hall resistance measured at various temperatures. Inset: The charge carrier density plotted as a function of temperature.

Figure 5: Normalized resistivity as a function of temperature of an 85 nm $Ta_3Sb$ film showing the superconducting transition at various applied magnetic fields. Inset: The superconducting phase diagram showing the upper critical field as a function of temperature.

Figure 6: (a) Schematic of the harmonic Hall measurement. (b) Field dependence of the first and second harmonic Hall voltages of a $Ta_3Sb$(9.5nm)/CFB(1nm) device measured at various

temperatures. The solid lines are curve fits. (c) The first harmonic Hall voltage as a function of the azimuthal angle for the same device measured with an in-plane field. (d) Temperature dependence of the DL and FL SOT efficiencies.

Figure 7: (a) ST-FMR spectra of a $Ta_3Sb$(9.5nm)/Py(5nm) sample measured at various frequencies. (b) The resonant frequency as a function of the magnetic field. The solid line is a fit using the Kittel formula. (c) Half linewidth of the ST-FMR spectra as a function of the resonant frequency. The solid line is a linear fit.

Figure 8: (a) Calculated band structure of $Ta_3Sb$ with SOC included. The dot-dash line indicates the Fermi level when 20% of Sb is replaced by Ta. (b) The change in $E_F$ plotted as a function the occupancy number for $Ta_3Sb$.

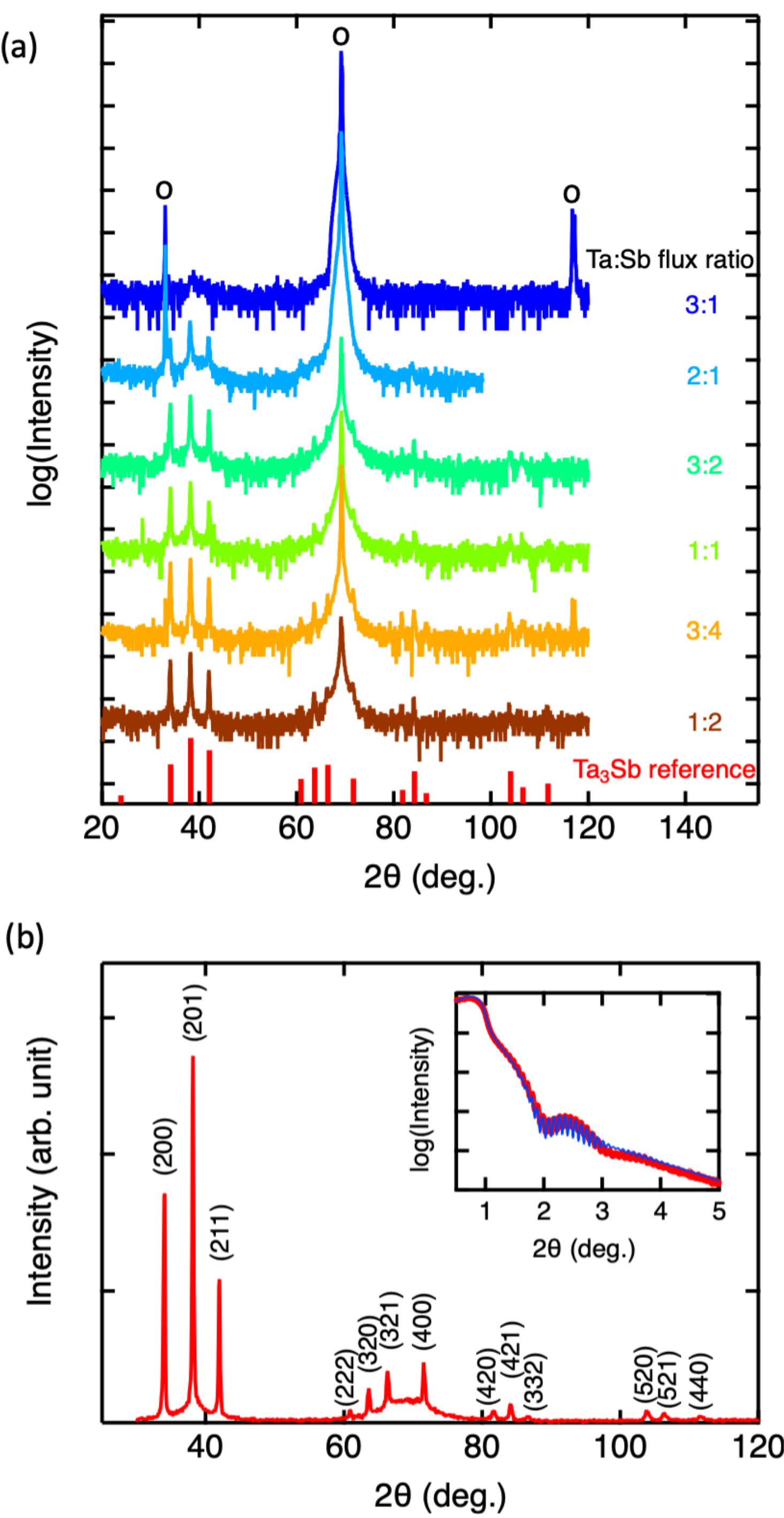

Figure 1:(a) XRD patterns of Ta-Sb films sputter-grown at various Ta:Sb flux ratios. The patterns are shifted vertically for clarity. The XRD scattering vector is aligned along the surface normal. The diffraction peaks marked with the symbol "o" are due to the silicon substrate. (b) XRD pattern of an 85 nm $Ta_3Sb$ film. Inset: XRR data and fit. The slower modulation in intensity is due to the MgO capping layer.

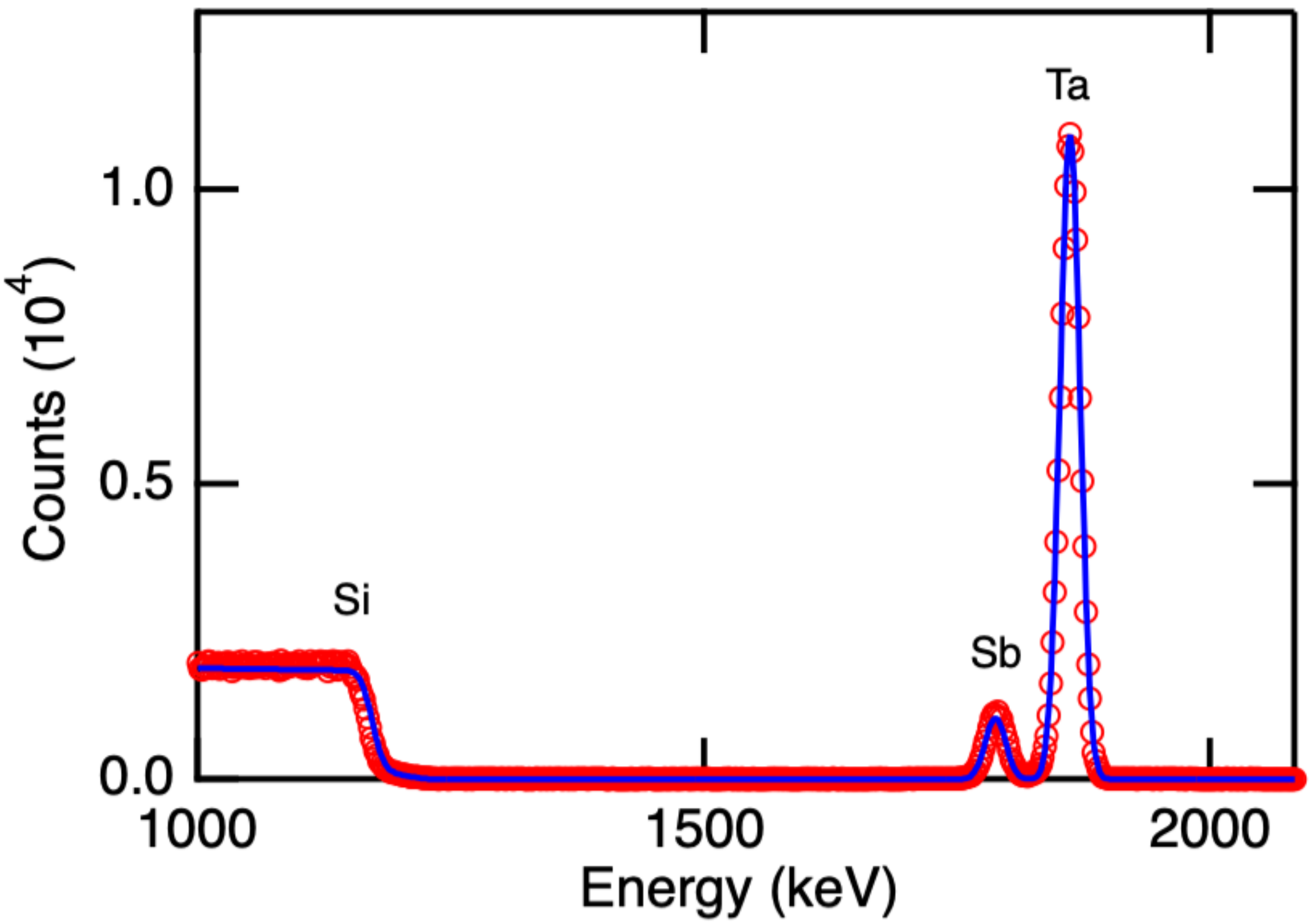

Figure 2:RBS data and fit of a 5.4 nm $Ta_3Sb$ film.

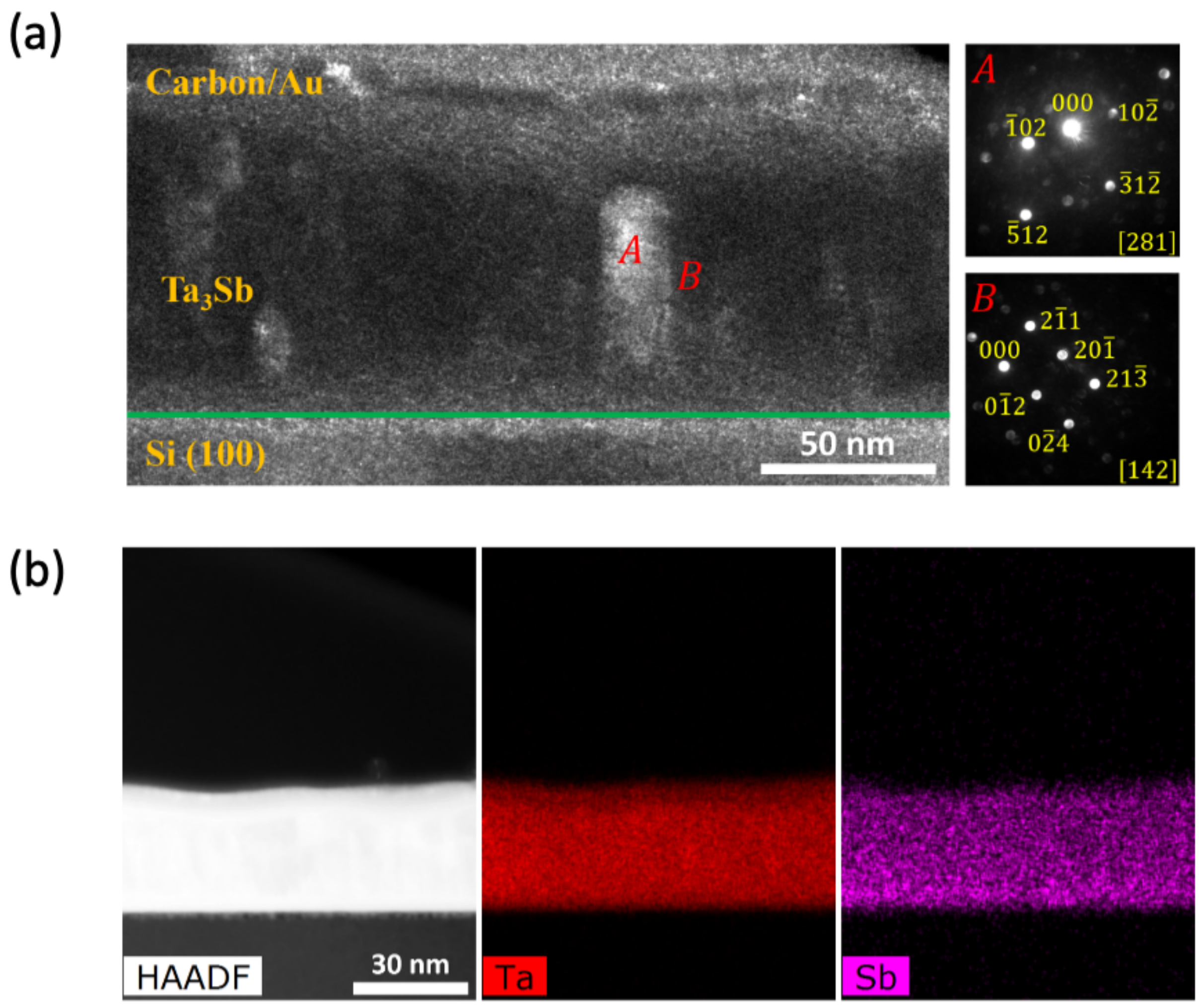

Figure 3:(a) Dark-field cross-sectional TEM image of an 85 nm $Ta_3Sb$ film. Nanobeam electron diffraction patterns of selected areas A and B are shown. (b) HAADF image and corresponding EDS maps of Ta and Sb showing the elemental distributions.

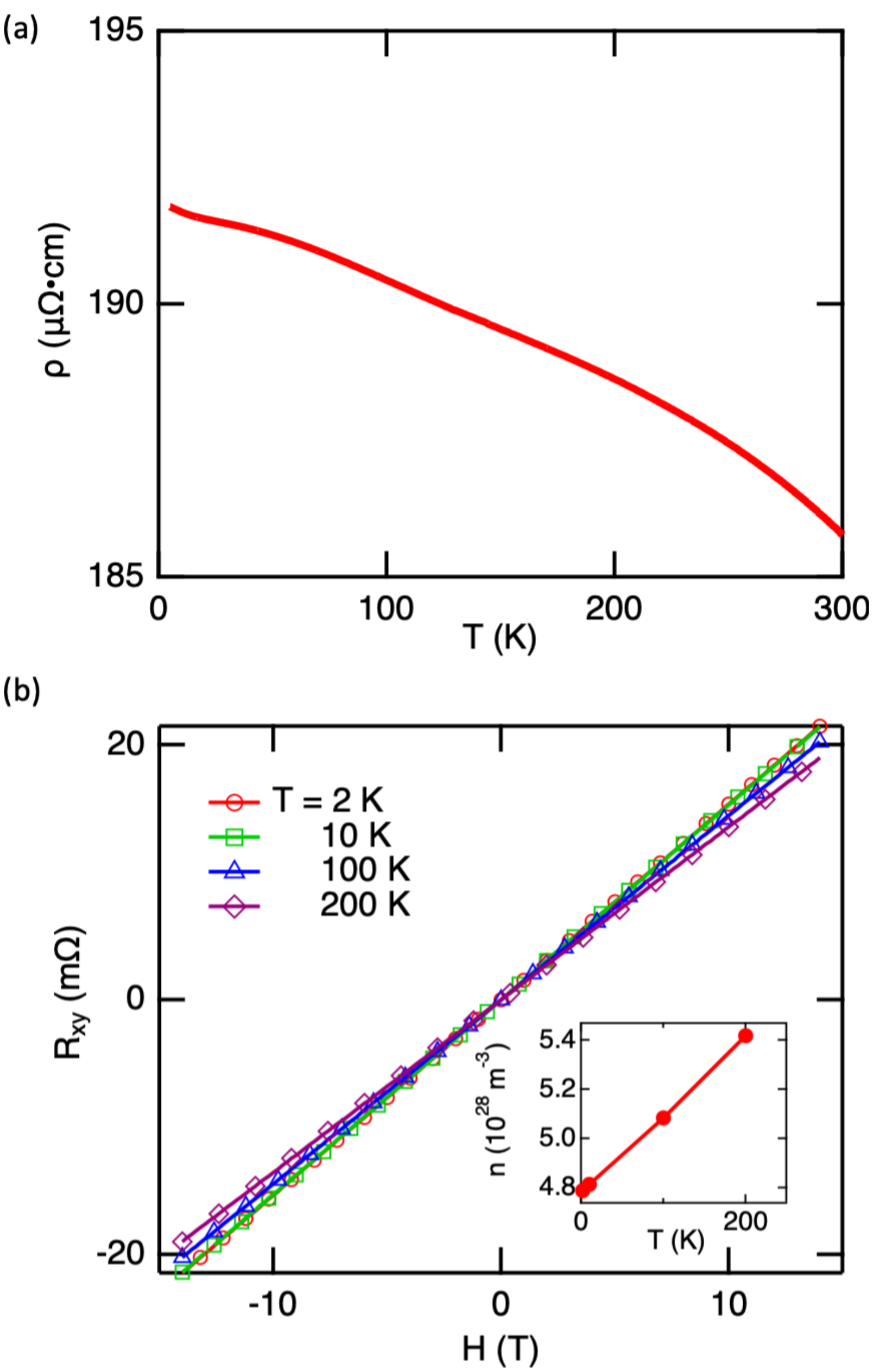

Figure 4:(a) Electrical resistivity of an 85 nm $Ta_3Sb$ film plotted as a function of temperature. (b) Hall resistance measured at various temperatures. Inset: The charge carrier density plotted as a function of temperature.

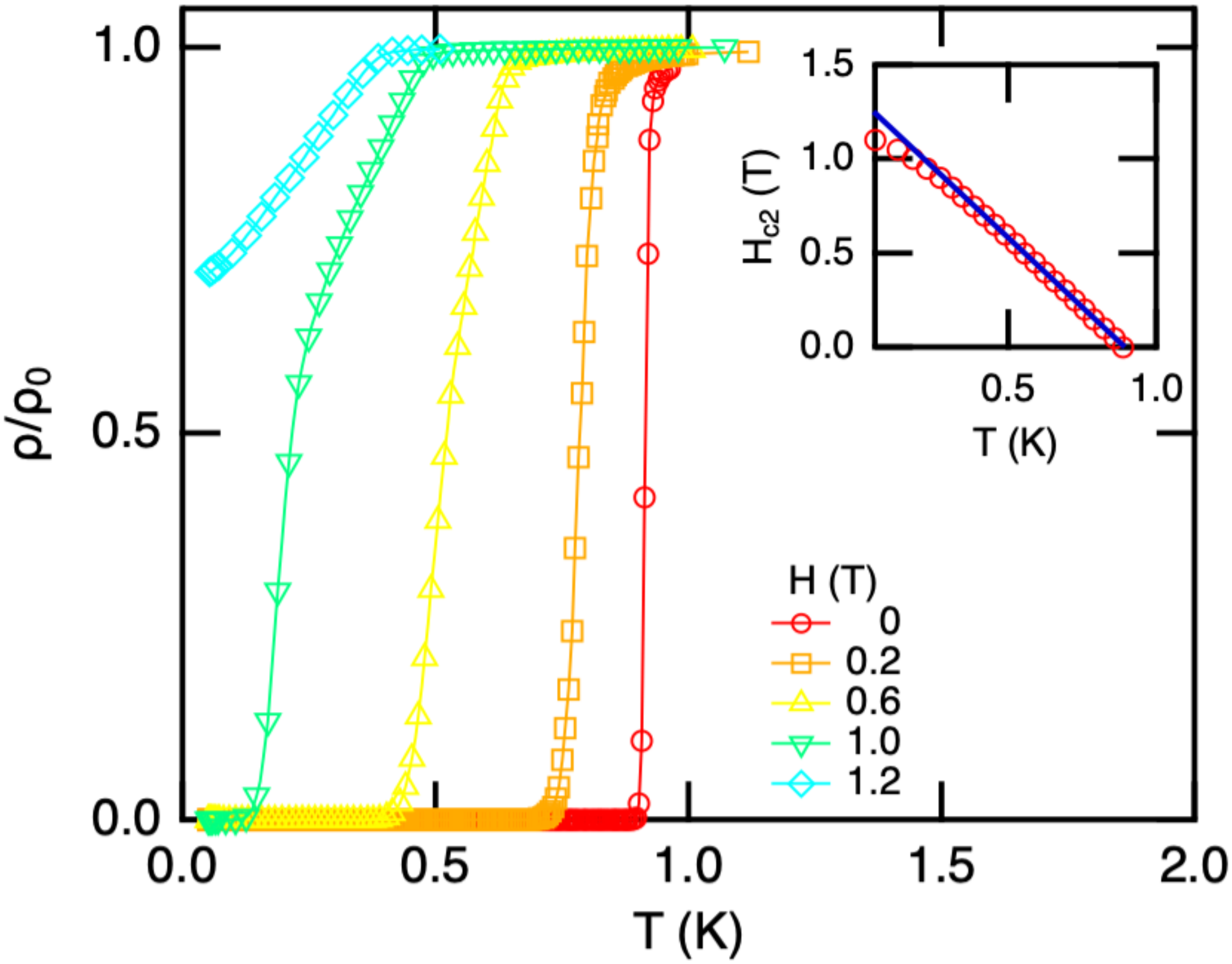

Figure 5:Normalized resistivity as a function of temperature of an 85 nm $Ta_3Sb$ film showing the superconducting transition at various applied magnetic fields. Inset: The superconducting phase diagram showing the upper critical field as a function of temperature.

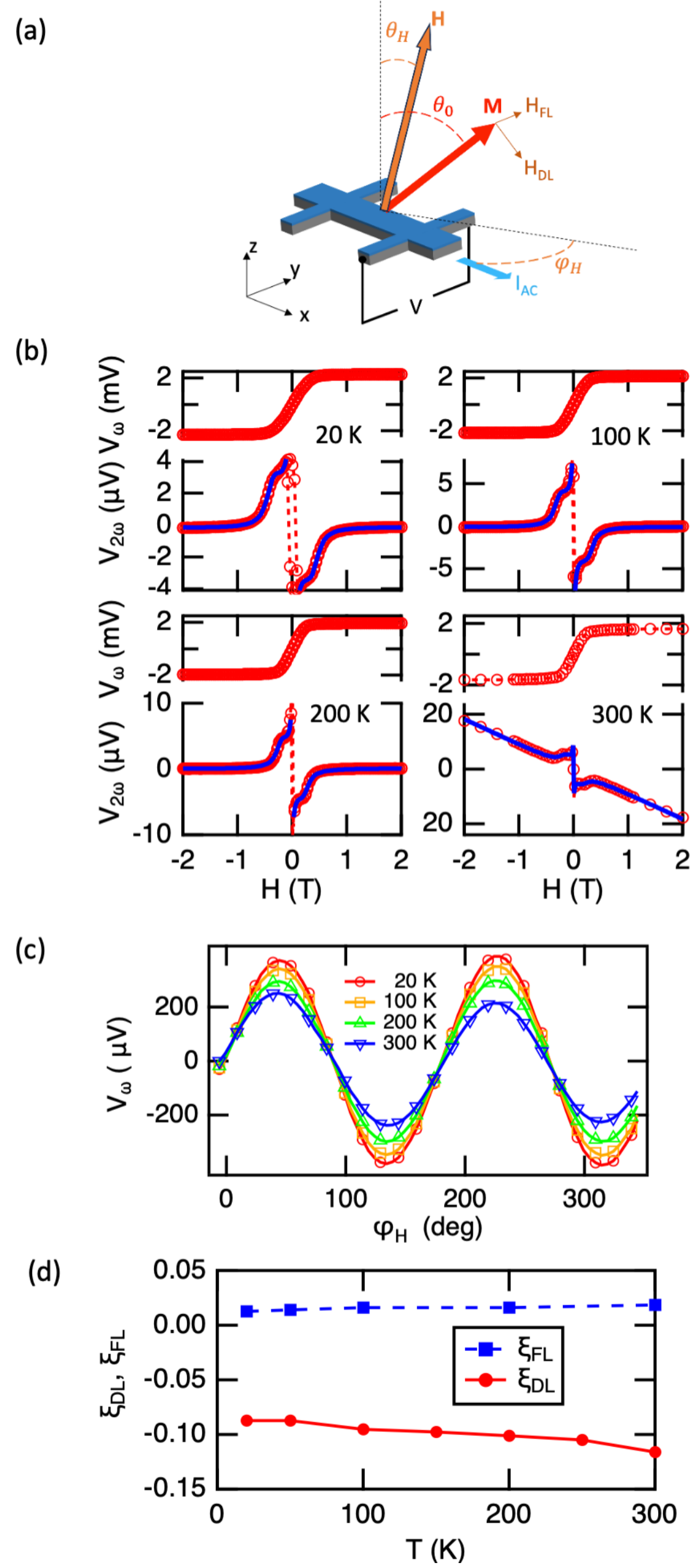

Figure 6:(a) Schematic of the harmonic Hall measurement. (b) Field dependence of the first and second harmonic Hall voltages of a $Ta_3Sb$(9.5nm)/CFB(1nm) device measured at various temperatures. The solid lines are curve fits. (c) The first harmonic Hall voltage as a function of the azimuthal angle for the same device measured with an in-plane field. (d) Temperature dependence of the DL and FL SOT efficiencies.

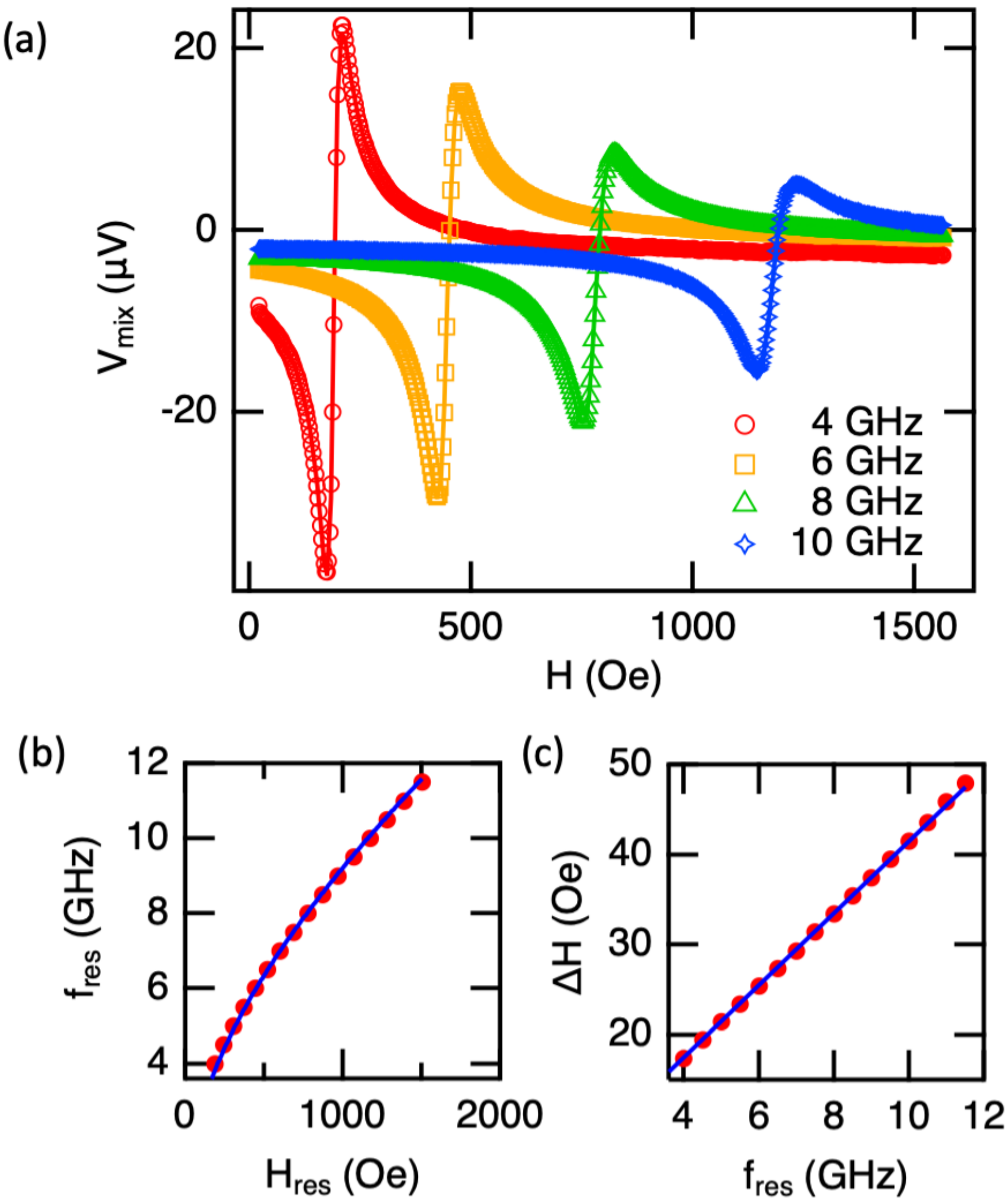

Figure 7:(a) ST-FMR spectra of a $Ta_3Sb$(9.5nm)/Py(5nm) sample measured at various frequencies. (b) The resonant frequency as a function of the magnetic field. The solid line is a fit using the Kittel formula. (c) Half linewidth of the ST-FMR spectra as a function of the resonant frequency. The solid line is a linear fit.

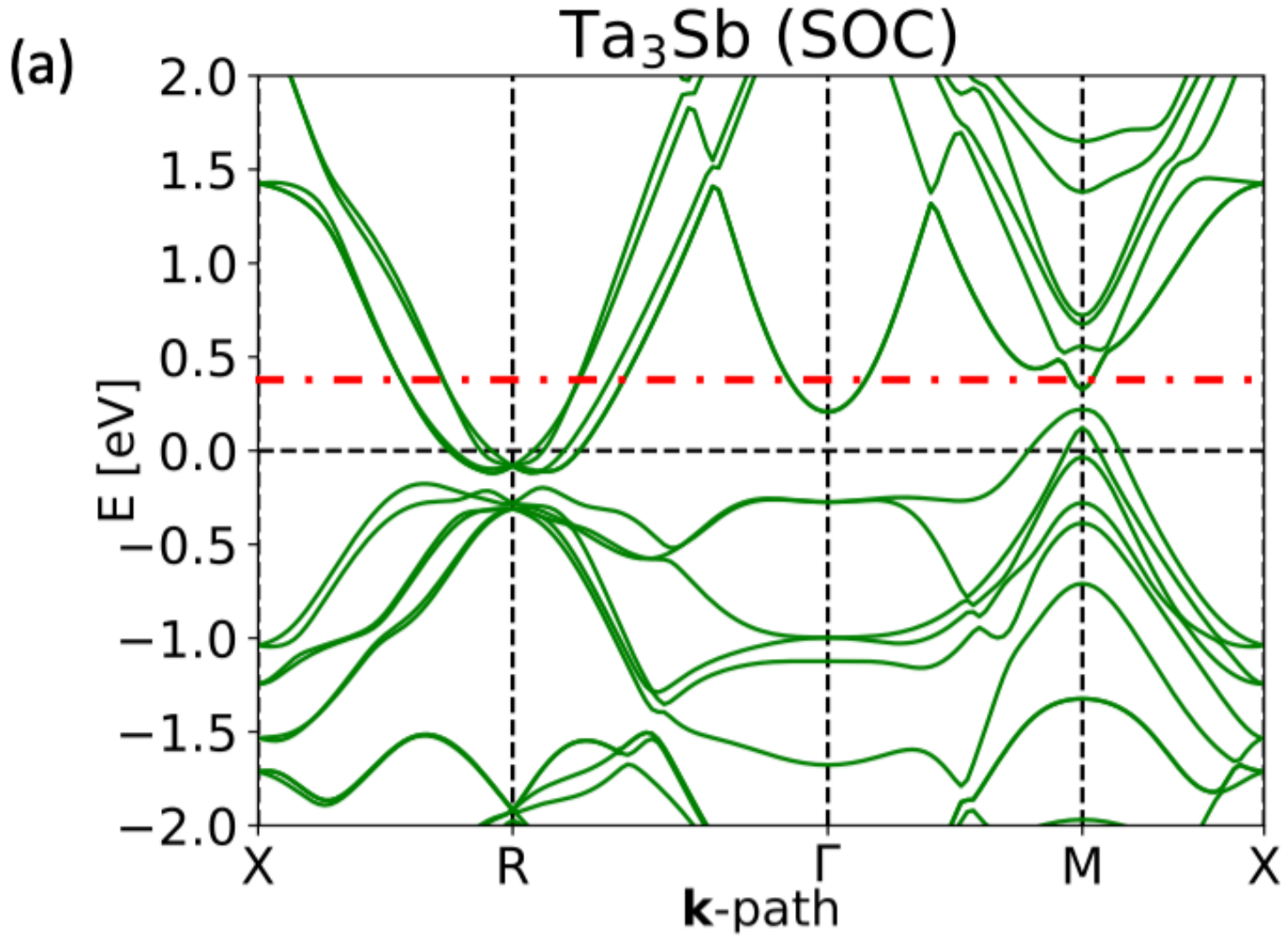

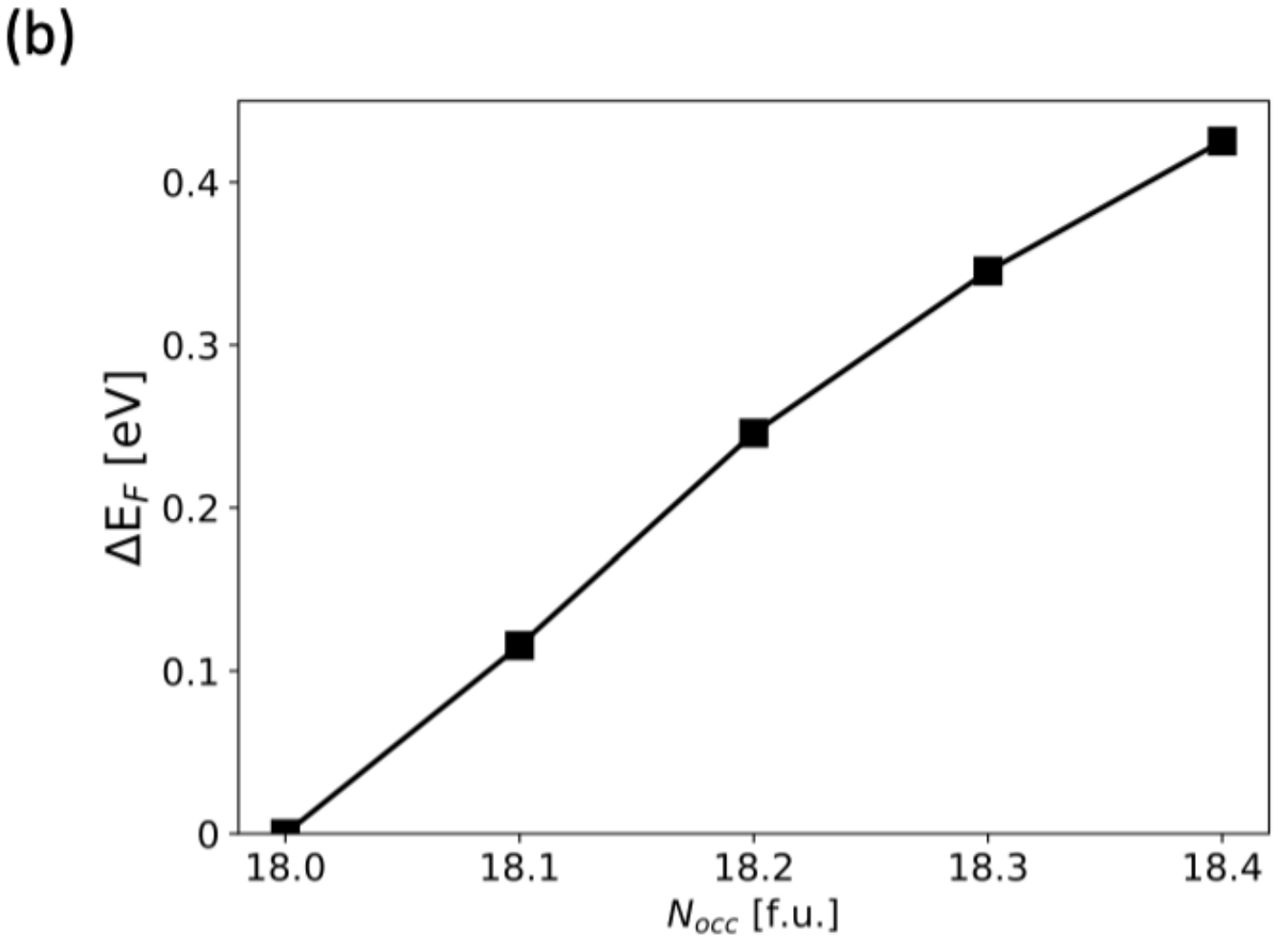

Figure 8:(a) Calculated band structure of $Ta_3Sb$ with SOC included. The dot-dash line indicates the Fermi level when 20% of Sb is replaced by Ta. (b) The change in $E_F$ plotted as a function the occupancy number for $Ta_3Sb$.